\documentclass[preprint]{aastex}

\slugcomment{To appear in Publications of the Astronomical Society of the Pacific}

\shorttitle{Diffracto-Astrometry measurements.}
\shortauthors{Ruelas-Mayorga et al.}

\begin{document}

\title{Diffracto-Astrometry measurements: accuracy of the measuring algorithm.}

\author{A. Ruelas-Mayorga\altaffilmark{1},
        L.J. S\'anchez\altaffilmark{1},
        J. Olivares\altaffilmark{1},
        C. Allen\altaffilmark{1},
        A. Poveda\altaffilmark{1},
        R. Costero\altaffilmark{1}}

\and
\author{A. Nigoche-Netro\altaffilmark{2}}

\altaffiltext{1}{Instituto de
Astronom\'{\i}a, Universidad Nacional Aut\'onoma de M\'exico, Apdo. Postal 70-264, 04510,
M\'exico D.F., M\'exico.}
\altaffiltext{2}{Instituto de Astrof\'{\i}sica de
Andaluc\'{\i}a (IAA), Glorieta de la Astronom\'{\i}a s/n, 18008, Granada, Spain.}

\begin{abstract}
We present a theoretical analysis of the measuring algorithm we
use when applying the Diffracto-Astrometry technique to Hubble
Space Telescope Wide Field Planetary Camera 2 (WFPC2) saturated
stellar images. Theoretical Point Spread Functions
(PSFs) were generated using the Tiny Tim software for the four
CCDs in the WFPC2 and for some of the available filters.

These images were then measured with our Diffracto-Astrometry
measuring algorithm using only their diffraction pattern, and
positions for the simulated PSFs on each generated CCD-frame were
obtained. The measuring algorithm recovers the original positions
reasonably well ($\pm 0.1\ \rm{to} \ \pm 0.4$ pixels). However,
slight deviations from the original values are observed. These
also vary with position over the entire surface of the CCD. We
adjust the difference between the real and the measured position
with a quadratic function of the coordinates. The transformation
coefficients also present a slight correlation with the filter
effective wavelength. Application of these transformation coefficients
allows us to determine the position of a stellar image with a
precision of a few hundredths of a pixel.
\end{abstract}


\section{Introduction}
\label{sec:introduction}

In the past there have been attempts to perform
astrometry as well as photometry on saturated sources in Hubble Space Telescope (HST) images. For example, Golimowski et al. (1995) used the diffraction spikes to
do astrometry with a precision of 0.5 pixels, Gilliland (2007) proposed a way to calibrate the use of the diffraction spikes to locate
 the photocentre of a saturated stellar image. Both Gilliland (1994) and Ma\'{\i}z-Apellan\'{\i}z (2002) have
developed somewhat different methods for obtaining photometry of saturated sources on HST images. The results appear promising. However, a definitive  technique for obtaining high precision astrometric measurements does not seem to be available.

Diffracto-Astrometry is a technique developed with the intention
of making high precision measurements of absolute and relative
stellar positions of saturated stellar images
\citep{sanchezetal2008}.  To
establish a position for a saturated stellar image,
Diffracto-Astrometry utilises in principle both the diffraction spikes and the
diffraction rings on the PSFs  \citep{sanchezetal2011a},
\citep{olivareslarim2011}.
So far, we have only explored the use of the diffraction spikes to determine the photocentre of
saturated stars on HST Wide Field Planetary
Camera 2 (WFPC2) images of the Orion Trapezium (OT) see \citep{sanchezetal2011b}.

To investigate the precision with which the measuring algorithm is able to recover the positions of saturated stars on archival observations, we study in this paper the measuring accuracy of the algorithm on WFPC2 simulated saturated images.

In \S2 we give a brief presentation of the
Diffracto-Astrometry technique and a detailed description of the measuring algorithm; in \S3 we determine the accuracy
of the measuring algorithm using a series of theoretical PSFs for some filters of the WFPC2; and \S4 gives
our conclusions.

\section{Difracto-Astrometry Technique and the Measuring Algorithm}
\label{sec:measuringalgorithm}

Diffracto-Astrometry is a methodology developed by us with the aim
of measuring absolute and relative stellar positions, and
displacements on diffraction-limited images such as archival HST
 and also Adaptive Optics (AO) images. Application of this
methodology is independent of whether or not the sources
are saturated. Successful application of this methodology to saturated images
enables us to extract valuable astrometric data from HST archival material.

The application
of the Diffracto-Astrometry techniques to the measurement of relative stellar positions
on archival HST images requires an algorithm to ensure minimum systematic errors in the
results (\citep{sanchezetal2008}, \citep{sanchezetal2011a}, \citep{sanchezetal2011b}).

There exist a number of systematic errors associated with the WFPC2 measurements which have been extensively analysed in the past. The most important are: geometric distortion (Kozhurina-Platais, et al. (2003) and Anderson \& King (2003)), the 34th row error (Anderson \& King (1999)), and Charge Transfer Efficiency (CTE) (Holtzman et al. (1995), Stetson (1998), Whitmore at al. (1999), and Dolphin (2000)).

In our analysis of the OT images, we correct for geometric distortion and the 34th row error using the formulae given in the literature mentioned above. Since the maximum charge loss due to the CTE problem amounts to a maximum $\sim 2 \%$ for bright sources, CTE problems are negligible when applying the Diffracto-Astrometry technique.

In what follows we enumerate the steps taken by the measuring algorithm in the determination of
the photocentre of an astronomical object.

\begin{enumerate}

\item First, we select standard pipeline-calibrated WPCF2 archival files (*.c0f.fits) from the Multimission Archive
at STScI (MAST) according
to criteria regarding astronomical object, date of acquisition, filter, exposure time and CCD (PC or
WF chip). The specific criteria for each one of these categories depends on the particular
astronomical aims (e.g. measurement of the separation between multiple stellar systems components at different dates, orbit, and proper motion determinations, etc).

To minimise possible effects of geometrical
distortions, the selected images must, preferably, have the astronomical objects as close to the centre of
the CCD as possible.

\item As this technique may be used for stellar images that may be saturated or present bleeding, it is necessary to
create a mask that identifies the saturated pixels and puts their value at a level which is determined interactively
for each image.

\item The angle formed by the CCD rows and the diffraction spikes of the brightest component is measured.

\item A straight line is fitted to each pair of
spikes and its intersection is taken as a first approximation of the pixel corresponding to the stellar photocentre.
The fitting of each straight line is achieved by means of an iterative method.

\item This pixel is used as the intersection point of a pair of perpendicular straight lines which are used as guidelines for
intensity cuts along the spikes. {\bf We assume that our initial guess locates the real photocentre within a square 3x3 pixels in size.}

\item  The faintest, most external point on any one of the diffraction spikes is chosen manually. This point determines the maximum spatial extension, measured from the centre, to which the measuring algorithm is applied to analyse the stellar image.

\item  Depending on the intensity modulating pattern of the spikes the algorithm chooses to effect regularly spaced perpendicular cuts across the maximum spatial extension of the diffraction spikes. The maximum spatial extension is variable (from a few to tens of pixels) and depends on the brightness of the diffraction spikes which, in turn, depends {\bf on the total amount of energy collected. It also depends on whether or not the spike image overlaps other luminous features (e.g. bleeding, other spikes or stars, etc).}

\item To each brightness-cut the algorithm adjusts a Gaussian curve by the Levenberg-Marquardt method \citep{more78}. The central point of this Gaussian is taken as the position of the spike on that cut.  The obtained central positions and their uncertainties for each cut are corrected for geometrical distortion by the Kozhurina-Platais et al. (2003) method which is an extension of the Anderson \& King (2003) geometric solution. The corrected central positions are used to generate a least-squares straight line which we consider as the best position for the diffraction spike. The intersection of the two least-squares lines fitted to both pairs of spikes provides a statistically significant position for the stellar photo-centre.

\item  Steps 5 to 8 are now repeated over the 8 surrounding pixels of the point determined in the fourth step, and a new statistically significant central position is determined for each case.
\item These positions are corrected for the 34th row error following Anderson \& King (1999)

\item Each one of the nine positions are finally corrected for the measuring algorithm intrinsic errors. The determination of these intrinsic errors is developed in \S \ref{sec:psf}.

\item The position which the measuring algorithm provides as the definitive position of a stellar image corresponds to the average of the 9 central positions determined in step 9. The uncertainty of this result corresponds to the standard deviation.

\end{enumerate}

\section{Theoretical Point Spread Functions (PSFs) used to determine the accuracy of the measuring algorithm}
\label{sec:psf}

It is well known that the shape of the WFPC2 PSF varies across the surface of the CCDs, as well as with the relative orientation of the diffraction spikes with respect to the CCDs' rows. Given that our algorithm is based on the determination of the spike positions to find stellar photocentres, these variations of shape and orientation affect the measurements. In this section,
 we study the intrinsic errors and determine the accuracy of the measuring algorithm mentioned in \S \ref{sec:measuringalgorithm}. To do this, we generate, using the Tiny Tim software \citep{krist1993, krist1995, kristhook2004, hook2008}, theoretical PSFs for 9 different, regularly distributed positions on CCD 2 (see Table~\ref{tab:table1}, and for 16 positions on CCDs 1, 3, and 4 of the WFPC2 (see Table~\ref{tab:table3}). We choose a different spatial sampling to investigate whether this has any influence on the results. The Tiny Tim software proves to be particularly useful since it simulates the proper obscuration function and determines field dependent aberrations across the surface of the different CCDs. Moreover, it simulates various effects such as including changes in the shapes of the diffraction rings, "squishing" of the PSF core, and alteration of the banding patterns in the diffraction spikes (\citep{kristhook2004}).

\begin{figure}[tbp]
\plotone{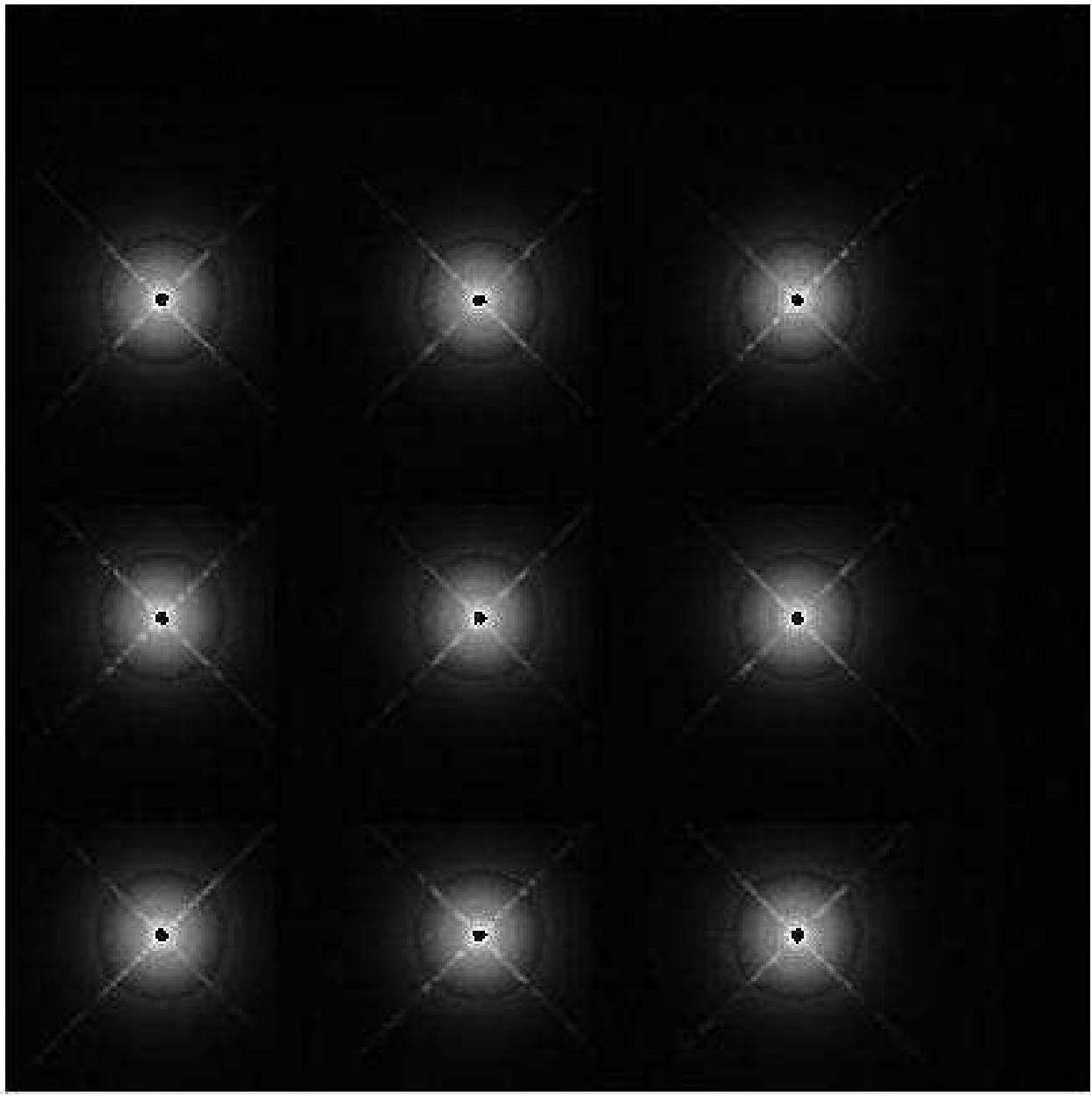} \caption{Theoretically generated Point Spread
Functions for CCD 2 and filter F1042M of the WFPC2. Due to the broad halo component anomaly specific to this filter, the diffraction features appear enhanced.} \label{fig:PSFFORF1042M}
\end{figure}

 We used the majority of the filters available with the WFPC2 only for the CCD 2 simulations and the size of each PSF was chosen to be equal to the maximum size which the Tiny Tim software is able to generate for each different filter (see Table~\ref{tab:table2}). Appropriate background and saturation levels were added to each generated image. In Fig.~\ref{fig:PSFFORF1042M} we show one of the images generated for
CCD 2 for filter F1042M. We chose to show this figure because the
diffraction features are much more noticeable for longer wavelength filters; in particular, for this filter there is a broad halo component anomaly which enhances the diffraction features. Since our measuring algorithm is based on the diffraction spikes, the broad halo anomaly of filter F1042M seems to have no effect on the results.

\begin{figure}[tbp]
\plotone{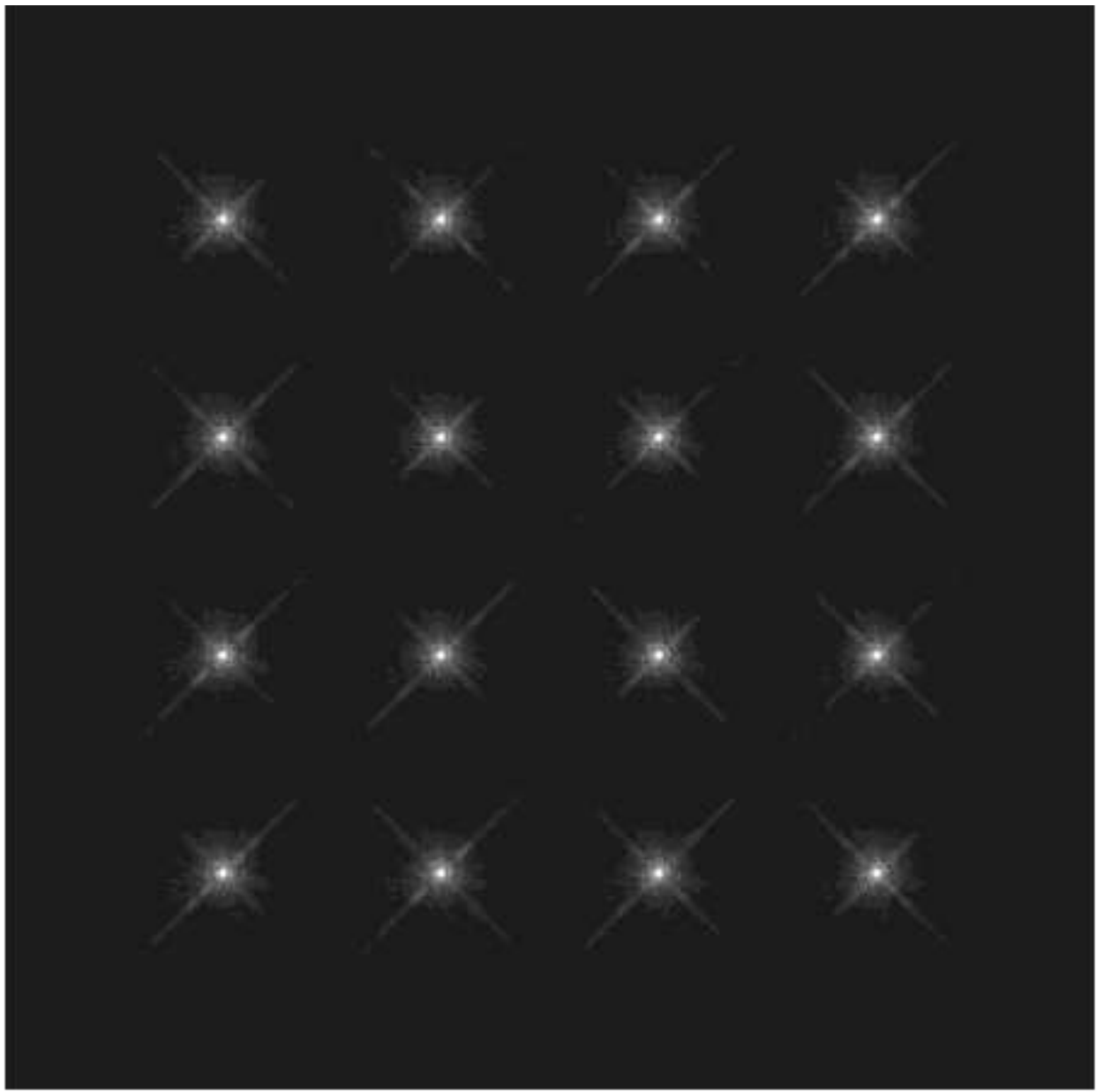}
\caption{Theoretically generated Point Spread
Functions for CCD 1 and filter F814W of the WFPC2.} \label{fig:CHIP1F814W}
\end{figure}

For CCDs 1, 3, and 4 we used fewer filters (see Table~\ref{tab:table4}). Fig.~\ref{fig:CHIP1F814W} shows the image generated for CCD 1 in filter
F814W. Again, we present this image because the diffraction patterns are better noticed in the longer
wavelength regime. The
simulated PSFs images for CCDs 3 and 4 were similar to those presented for CCD 1.

For each one of the theoretical PSFs across the field we determined the photocentre applying the algorithm described in \S \ref{sec:measuringalgorithm}. The determinations of the PSF photocentres differ slightly from the theoretical positions.

{\bf After having experimented with different polynomial functions, we chose to fit the differences measured for each positional coordinate with the following polynomials which represent a compromise between lower residual values and polynomials that did not produce too many folds in the fitting surface. We also stayed away from higher degree polynomials since, as is well known, anything can be exactly fitted with a high enough degree function:}

\begin{eqnarray}
\mathrm{\Delta x}=\mathrm{measured-theoretical}=\mathrm{(AX)x+(BX)x^2+(CX)y+(DX)y^2+(EX)} \nonumber\\
\end{eqnarray}

and

\begin{eqnarray}
\mathrm{\Delta y}=\mathrm{measured-theoretical}=\mathrm{(AY)x+(BY)x^2+(CY)y+(EY)} \nonumber\\
\end{eqnarray}

The coefficients of equations (1) and (2) were found following a least-squares procedure. Note that in equation (2) coefficient DY does not appear since a quadratic term in y was not necessary. The values of the coefficients and their peak-to-peak errors for CCD 2 are given in Tables~\ref{tab:table5} and ~\ref{tab:table6} for the difference in X and Y, respectively, as functions of the filters used.

The values for the transformation coefficients for CCDs 1, 3, and 4 and their peak-to-peak errors are given in Table~\ref{tab:table7} (for the X differences) and in Table~\ref{tab:table8} (for the Y differences).

\begin{figure}[tbp]
\plotone{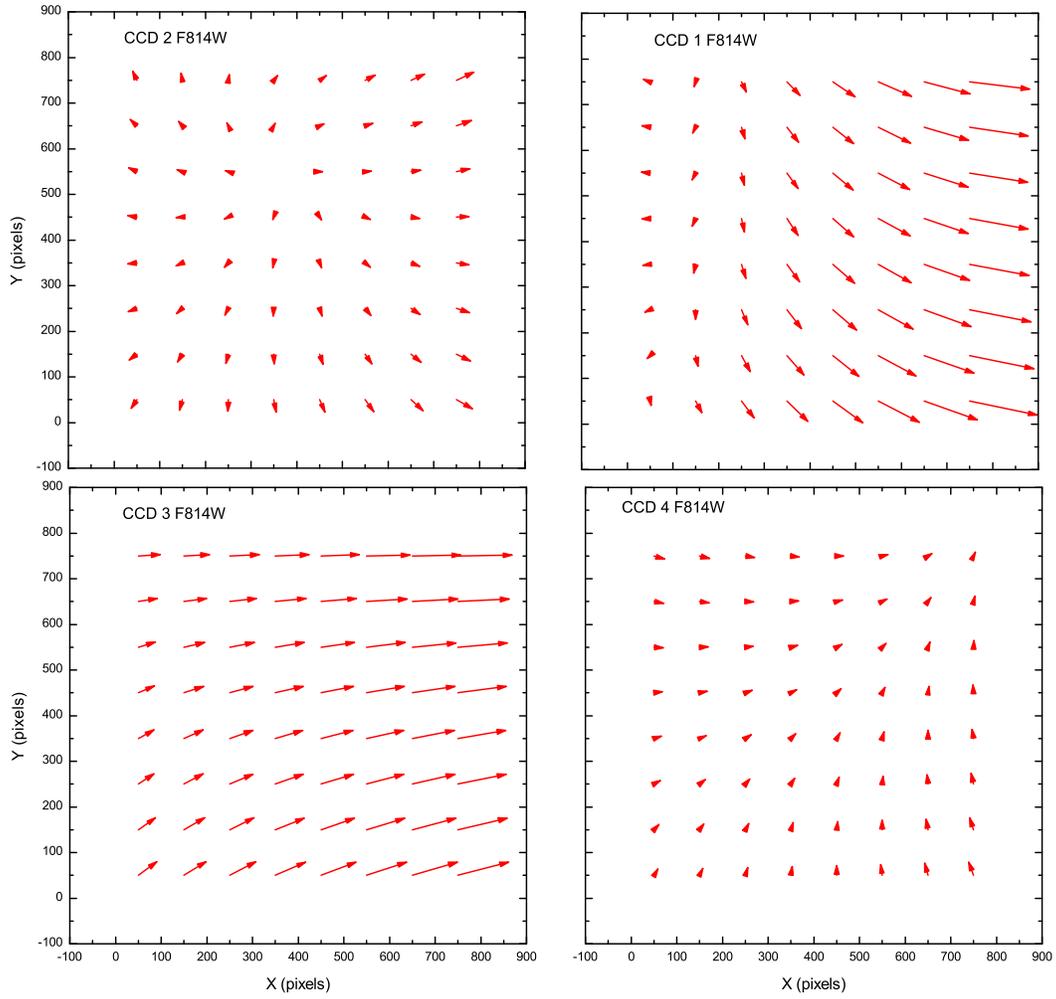}
\caption{Vector diagrams of the differences between the measured and the real positions accross the surface of
the WFPC2 CCDs for filter F814W. To make the differences visible, each vector is magnified by 100.} \label{fig:DEVIATIONS}
\end{figure}

\begin{figure}[tbp]
\plotone{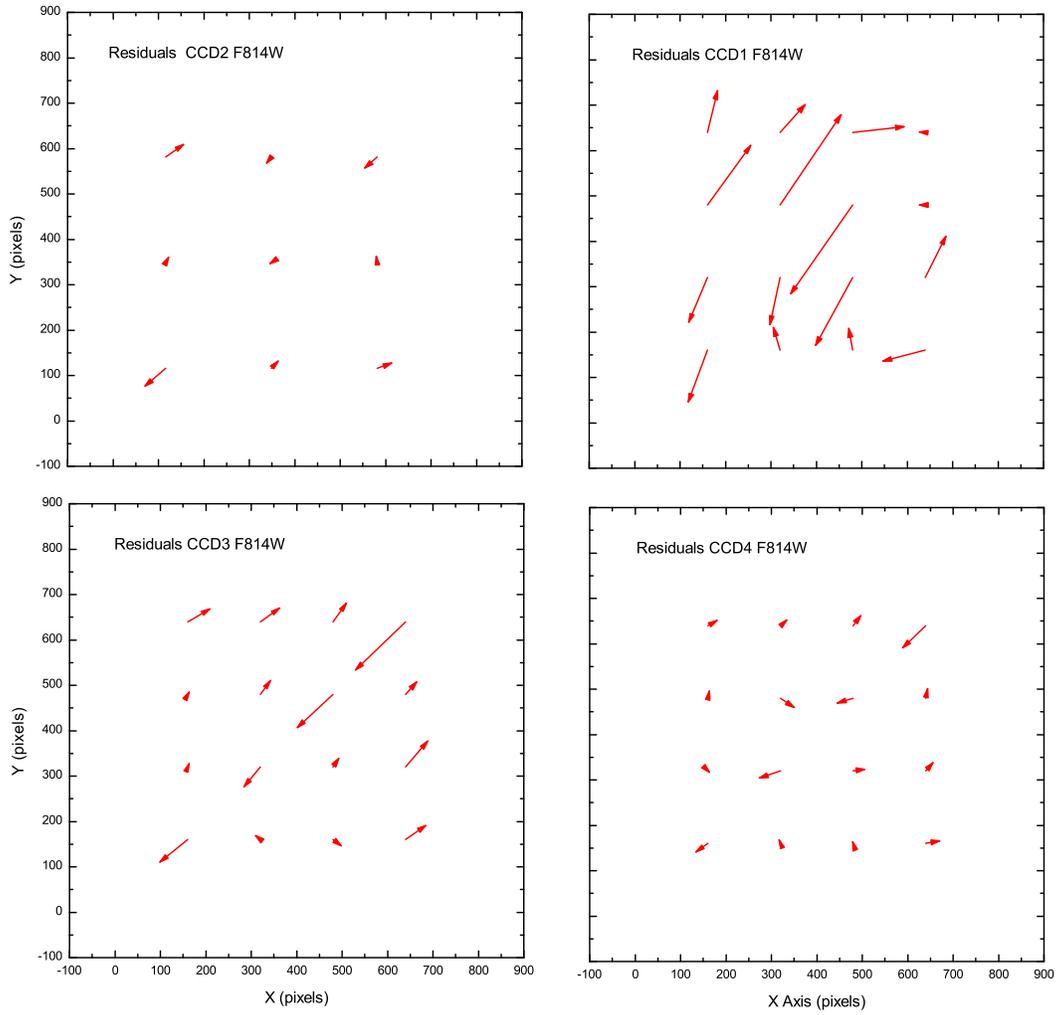}
\caption{Vector diagrams of the residuals between the measured and the real positions accross the surface of
the WFPC2 CCDs for filter F814W. To make the residuals visible, each vector is magnified by 500.} \label{fig:RESIDUALS}
\end{figure}

As an example, in Fig.~\ref{fig:DEVIATIONS} we show vector diagrams made for filter F814W and for the four CCDs in the WFPC2.  The vectors represent the
differences we find between the theoretical positions and those measured by our algorithm. The differences (in pixels) in X are contained in the range (0.11, 1.48), and in Y (0.02, 0.48). We also present, in Fig.~\ref{fig:RESIDUALS}, vector diagrams of the residuals between the  measured values and those produced by our fitting functions. The residual values (in pixels) in X are contained in the range (0.06, 0.27), and in Y (0.04, 0.40).

The peak-to-peak errors in the coefficients were calculated by fitting the coefficients to two extreme cases given by the uncertainties provided by the measuring algorithm for each measured position. If we choose to express the errors as the standard deviation, the peak-to-peak errors need to by multiplied by approximately $0.3$.

After having performed a statistical analysis of the measuring results for the 77 cases considered {\bf (41 filters for CCD2 and 12 filters each for CCDs 1, 3 and 4)}, we determined that  the Diffracto-Astrometry measuring algorithm is able to recover the PSF positions with an accuracy of $\pm 0.1-\pm 0.4$ pixels both in the X and Y coordinates. The correction to the measuring algorithm presented here should improve this accuracy. A statistical analysis of the residuals provides an indication of the accuracy with which the measuring algorithm is able to obtain the position of a PSF on the surface of the CCD. Table~\ref{tab:ACCURACY} gives the accuracy (standard deviation) for each CCD of the WFPC2. We see that in all cases the accuracy is improved, although for CCD 1 it is substantially larger than for the other CCDs. At present we are investigating the possible causes of this difference, as it might be due to limitations of the PSF modelling process, which includes: limited sampling, wavelength dependence of optical and detector responses, and field dependence of aberrations.

\begin{figure}[tbp]
\plotone{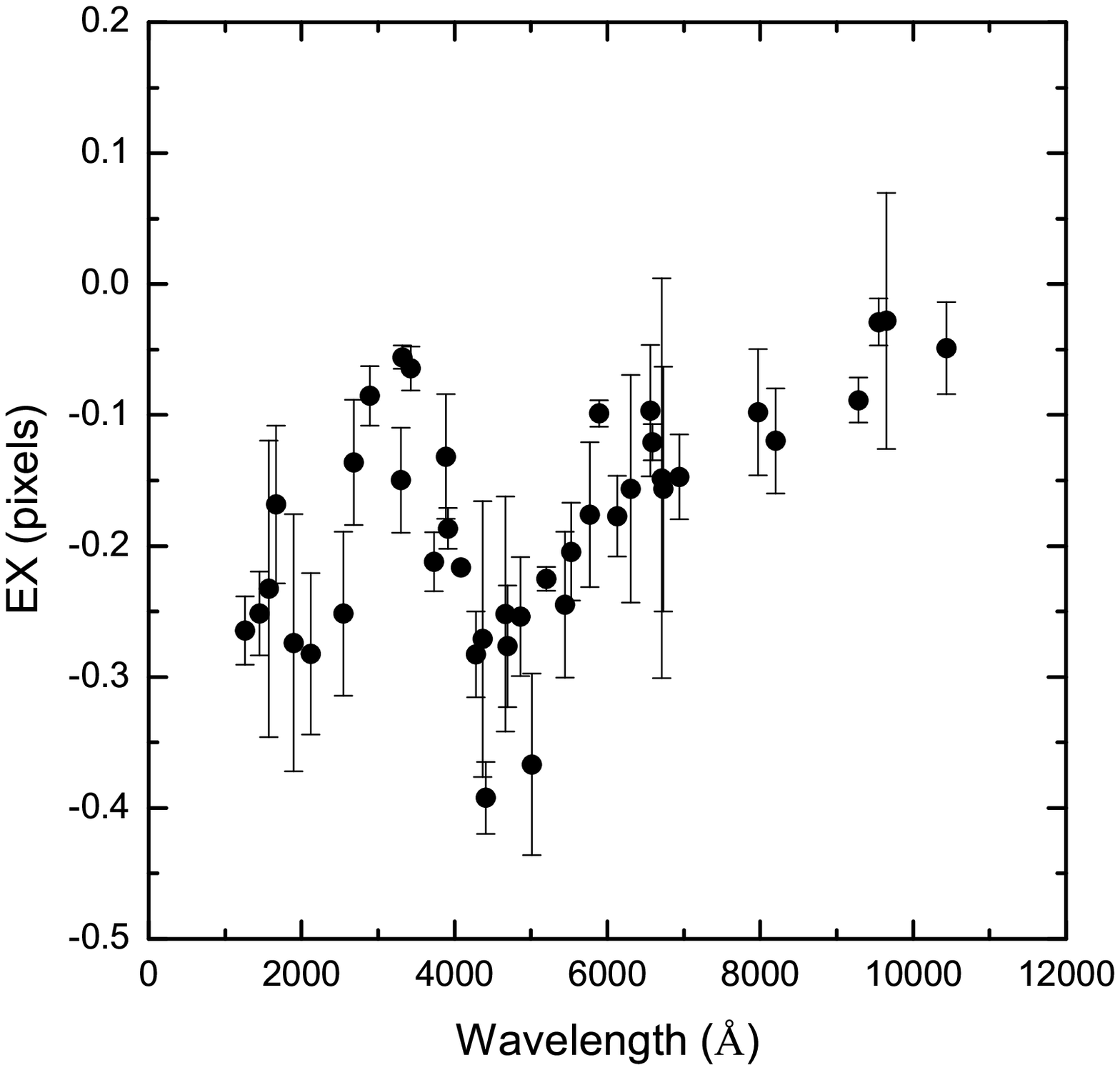}
\caption{Dependence of the coefficient EX with
wavelength for CCD 2 of the WFPC2.} \label{fig:EXvsLAMBDACHIP2}
\end{figure}

\begin{figure}[tbp]
\plotone{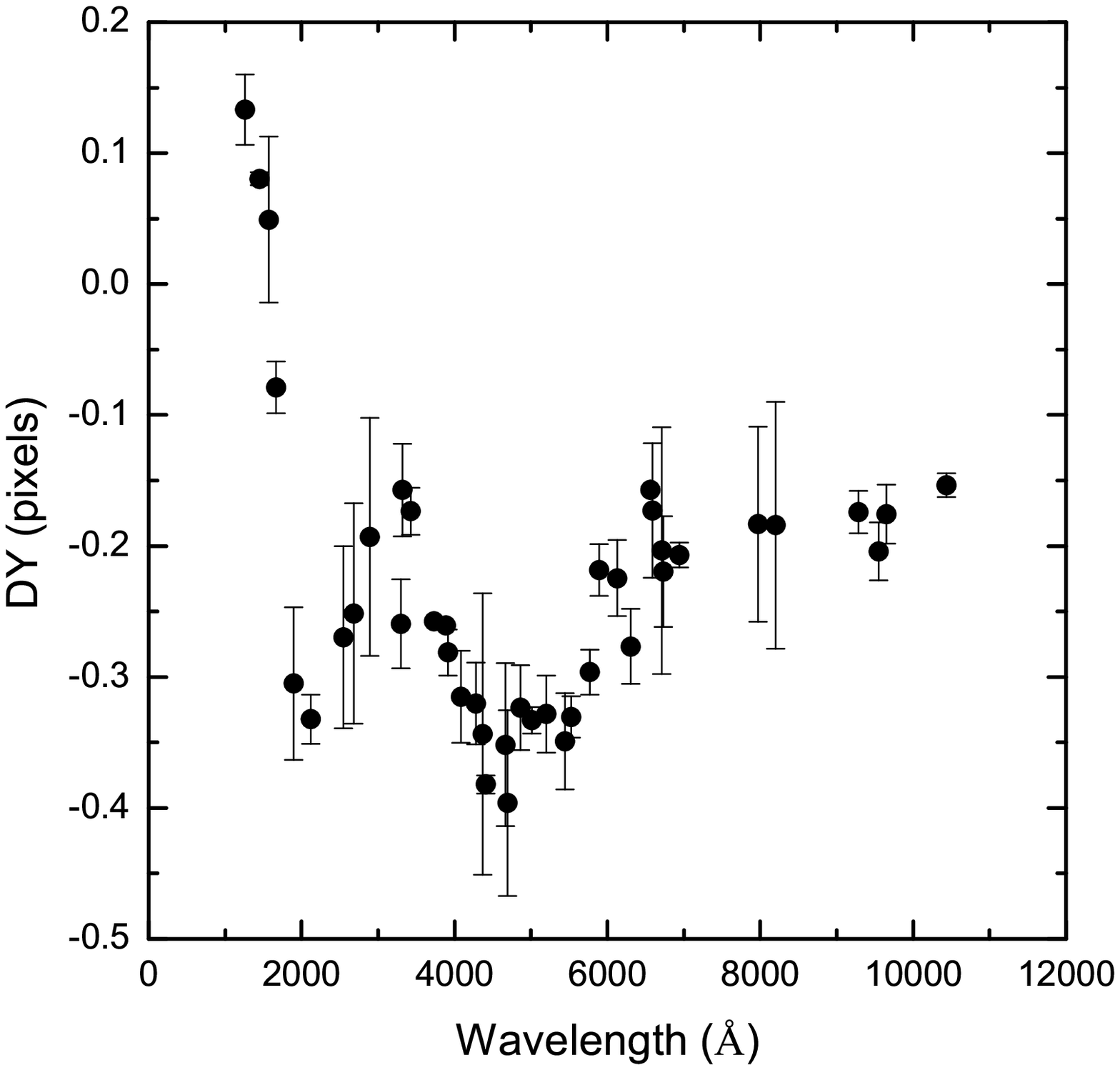}
\caption{Dependence of the coefficient DY with
wavelength for CCD 2 of the WFPC2.} \label{fig:DYvsLAMBDACHIP2}
\end{figure}

For CCD 2 for which we performed measurements for 41 different filters, there appears to be a slight dependence of the values of the coefficients of transformation on wavelength.
To illustrate this dependence we present a figure for the X and Y coefficients for which
the dependence is more noticeable. Fig~\ref{fig:EXvsLAMBDACHIP2}
shows the run of the value of the EX transformation coefficient for the X coordinate as a function of wavelength, while Fig.~\ref{fig:DYvsLAMBDACHIP2}
shows the run of the DY coefficient values, also as a function of wavelength.

{\bf In a recent paper \citep{bellini2011}, the authors find a different behaviour for very blue and very red stars which they attribute to a chromatic effect induced by fused-silica CCD windows within the optical system, which refract blue and red photons differently and have a sharp increase of the refractive index below $4000 \AA$, however, Tiny Tim does not simulate the effects of CCD windows, so the sharp decline of the value of DY in Fig.~\ref{fig:DYvsLAMBDACHIP2} cannot be explained in this way. We suspect that the variation of the size of the Tiny Tim simulated PSF with wavelength is, in part, a factor that causes the variation of the values of the transformation coefficients.}

To recover the astrometric information it is, therefore, very important to correct the measurements provided by our algorithm using the appropriate coefficients for each CCD and each different wavelength.

\section{Conclusions}
\label{sec:conclusions}

In this paper we present an analysis of the ability of the Diffracto-Astrometry measuring algorithm for finding the position of the centre of a saturated stellar image on HST images. As discussed in \S~\ref{sec:measuringalgorithm}, the algorithm assumes that the centre of a stellar image coincides with
the intersection point of two straight lines which are fitted, following a series of criteria, to the images of the diffraction spikes of a
stellar image on a HST exposure.

In \S~\ref{sec:psf} we present PSF simulations with the purpose of finding the accuracy of the measuring algorithm. We located these PSFs on the surface of the different CCDs and then compared their theoretical positions with those found by application of the algorithm. The differences between the measured
minus the real value are fitted to a quadratic function of the coordinates, and the transformation coefficients are
given in Tables ~\ref{tab:table5},~\ref{tab:table6},~\ref{tab:table7}, and ~\ref{tab:table8} for a large number of filters available in the WFPC2.
The Diffracto-Astrometry measuring algorithm alone recovers the stellar positions with a precision $\pm 0.1 \  \rm{to} \  \pm 0.4$ pixels, while after the application of these transformation equations the position is recovered with a precision of a few hundredths of a pixel of the real value (see Table~\ref{tab:ACCURACY}). For CCDs 2, 3, and 4 the accuracy varies between $0.02$ and $0.06$ pixels, while for CCD 1 it results in a larger interval (0.09--0.13 pix). At present we believe the difference to be caused by its different sampling rate.

We also found a slight dependence of the values of the transformation coefficients on the wavelength.

To ensure the best possible accuracy, when obtaining astrometric results, the application of the Diffracto-Astrometry technique requires the use of the appropriate coefficients, presented in this work, for each CCD and for each different wavelength.

Further advances and applications of these methods will be published elsewhere.

\acknowledgments

We are grateful to DGAPA--UNAM for financial support under project PAPIIT number IN109809. {\bf We acknowledge an anonymous referee for his/her numerous suggestions which greatly improved this work.}

\begin{table}[!h]
\caption{Simulated Theoretical PSFs Positions for CCD
2} \label{tab:table1}\centering

\begin{tabular}{|c|c|}
\hline \multicolumn{2}{|c|}{Relative Coordinates}           \\\hline
     x                &            y         \\\hline
     116              &          116         \\
     349              &          116         \\
     582              &          116         \\
     116              &          349         \\
     349              &          349         \\
     582              &          349         \\
     116              &          582         \\
     349              &          582         \\
     582              &          582         \\\hline
\end{tabular}
\end{table}

\begin{table}[!h]
\caption{Simulated Theoretical PSFs Positions for CCDs
 1, 3 and 4 of the WFPC2.} \label{tab:table3}\centering

\begin{tabular}{|c|c|}
\hline \multicolumn{2}{|c|}{Relative Coordinates}           \\
\hline
     x                &            y         \\\hline
     160              &           160        \\
     320              &           160        \\
     480              &           160        \\
     640              &           160        \\
     160              &           320        \\
     320              &           320        \\
     480              &           320        \\
     640              &           320        \\
     160              &           480        \\
     320              &           480        \\
     480              &           480        \\
     640              &           480        \\
     160              &           640        \\
     320              &           640        \\
     480              &           640        \\
     640              &           640        \\\hline

\end{tabular}
\end{table}

\begin{table}[!h]
\caption{Filters used for Theoretical PSFs simulations and PSF maximum diameter for the WFPC2 for CCD 2}
\label{tab:table2}\centering

\begin{tabular}{|c|c|c||c|c|c|}
\hline
   Filter       &   Effective          &   PSF diameter (\arcsec) &  Filter  & Effective       &   PSF diameter (\arcsec)                  \\
                &   Wavelength (\AA)   &                          &          & Wavelength (\AA)&                                       \\\hline
    F122M       &           1259       &     6.5                  & F469N    &           4694  &    25.7        \\
    F130LP      &           2681       &    10.3                  & F487N    &           4865  &    26.5        \\
    F157W       &           1570       &     6.5                  & F502N    &           5012  &    27.3         \\
    F160BW      &           1446       &     6.5                  & F547M    &           5446  &    27.5         \\
    F165LP      &           3301       &    10.3                  & F555W    &           5202  &    24.1         \\
    F170W       &           1666       &     6.9                  & F569W    &           5524  &    26.7         \\
    F185W       &           1899       &     6.9                  & F588N    &           5893  &    30.0         \\
    F218W       &           2117       &     9.9                  & F606W    &           5767  &    25.7         \\
    F255W       &           2545       &    12.9                  & F622W    &           6131  &    29.5         \\
    F300W       &           2892       &    12.9                  & F631N    &           6306  &    30.0         \\
                &                      &                          &          &                 &                 \\
    F336W       &           3317       &    16.3                  & F656N    &           6564  &    30.0         \\
    F343N       &           3427       &    18.3                  & F658N    &           6591  &    30.0         \\
    F375N       &           3732       &    20.3                  & F673N    &           6732  &    30.0         \\
    F380W       &           3912       &    18.1                  & F675W    &           6714  &    30.0         \\
    F390N       &           3888       &    21.1                  & F702W    &           6940  &    30.0         \\
    F410M       &           4086       &    21.5                  & F785LP   &           9283  &    30.0         \\
    F437N       &           4369       &    23.9                  & F791W    &           7969  &    30.0         \\
    F439W       &           4283       &    21.5                  & F814W    &           8203  &    30.0         \\
    F450W       &           4410       &    20.1                  & F850LP   &           9650  &    30.0         \\
    F467M       &           4663       &    24.7                  & F953N    &           9546  &    30.0         \\
                &                      &                          & F1042M   &          10437  &    30.0         \\\hline

\end{tabular}
\end{table}

\begin{table}[!h]
\caption{Filters used for Theoretical PSF simulations for WFPC2 CCDs 1, 3 and 4}
\label{tab:table4}\centering

\begin{tabular}{|c|c|}

                                                             \hline
       Filter                &      Effective             \\
                             &      Wavelength (\AA)       \\\hline
       F336W                 &             3317            \\
       F410M                 &             4086            \\
       F439W                 &             4283            \\
       F502N                 &             5012            \\
       F547M                 &             5446            \\
       F631N                 &             6306            \\
       F656N                 &             6564            \\
       F658N                 &             6591            \\
       F673N                 &             6732            \\
       F791W                 &             7969            \\
       F814W                 &             8203            \\
       F953N                 &             9546            \\\hline

\end{tabular}
\end{table}

\clearpage

\begin{deluxetable}{ccccccccccc}
\tabletypesize{\scriptsize} \rotate \tablecaption{Transformation
Coefficients for differences in X for CCD 2 of the
WFPC2.\label{tab:table5}}

\tablewidth{0pt}

\tablehead{ \colhead{Filter} & \colhead{AX} & \colhead{$\Delta$
AX} & \colhead{BX} & \colhead{$\Delta$ BX} & \colhead{CX} &
\colhead{$\Delta$ CX} & \colhead{DX}   &  \colhead{$\Delta$ DX}
&\colhead{EX} & \colhead{$\Delta$ EX} }

\startdata
F122M   &   5.6754E-04  &   7.4051E-04  &   -3.0700E-10 &   1.1187E-07  &   -1.0047E-03 &   3.8279E-04  &   1.0568E-06  &   6.0255E-07  &    -2.6460E-01    &   2.6083E-02 \\
F130LP  &   3.0497E-04  &   1.4972E-04  &   6.0562E-07  &   5.0247E-07  &   -6.0035E-04 &   1.7780E-04  &   6.5038E-07  &   4.1273E-07  &    -1.3617E-01    &   4.7912E-02  \\
F157W   &   9.4432E-04  &   2.0288E-04  &   -3.3988E-07 &   9.2023E-07  &   -1.3691E-03 &   4.2368E-04  &   1.6139E-06  &   5.8397E-07  &    -2.3269E-01    &   1.1324E-01  \\
F160BW  &   9.4833E-04  &   6.8330E-04  &   -3.6877E-07 &   1.6348E-07  &   -1.2723E-03 &   2.4316E-04  &   1.3996E-06  &   4.1869E-07  &    -2.5140E-01    &   3.2067E-02  \\
F165LP  &   3.4549E-04  &   8.2253E-04  &   5.6933E-07  &   5.0111E-07  &   -5.8386E-04 &   9.8663E-05  &   6.3205E-07  &   6.3733E-08  &    -1.4974E-01    &   4.0088E-02 \\
F170W   &   3.9367E-04  &   1.4344E-04  &   3.6158E-07  &   9.7997E-07  &   -7.2677E-04 &   2.2742E-04  &   5.7949E-07  &   4.6053E-07  &    -1.6835E-01    &   6.0364E-02  \\
F185W   &   8.9596E-04  &   1.4461E-05  &   -8.8723E-09 &   1.0273E-06  &   -9.8311E-04 &   3.1489E-04  &   1.0516E-06  &   4.7063E-07  &    -2.7394E-01    &   9.8201E-02  \\
F218W   &   7.1334E-04  &   7.5966E-04  &   1.6470E-07  &   3.6812E-07  &   -9.1635E-04 &   3.7352E-04  &   1.1771E-06  &   5.0001E-07  &    -2.8232E-01    &   6.1601E-02 \\
F255W   &   1.2543E-03  &   5.8644E-05  &   -3.8175E-07 &   8.8391E-07  &   -1.0872E-03 &   3.1794E-05  &   1.1629E-06  &   7.9636E-08  &    -2.5160E-01    &   6.2568E-02  \\
F300W   &   1.1182E-03  &   3.4309E-04  &   -2.0922E-07 &   2.9088E-07  &   -1.2859E-03 &   3.6894E-04  &   1.1874E-06  &   7.7867E-07  &    -8.5278E-02    &   2.2760E-02  \\
F336W   &   5.7441E-04  &   1.7049E-04  &   3.4412E-07  &   5.9134E-07  &   -8.4992E-04 &   3.4651E-04  &   6.3883E-07  &   4.9737E-07  &    -5.5832E-02    &   8.7825E-03 \\
F343N   &   5.5925E-04  &   2.8779E-04  &   2.9478E-07  &   3.5360E-07  &   -7.0045E-04 &   2.8551E-05  &   4.3588E-07  &   3.7454E-09  &    -6.4468E-02    &   1.6779E-02  \\
F375N   &   9.9401E-04  &   3.9191E-05  &   -2.3412E-07 &   7.3569E-07  &   -5.1349E-04 &   2.8513E-04  &   2.8376E-07  &   2.3964E-07  &    -2.1205E-01    &   2.2475E-02 \\
F380W   &   5.1561E-04  &   9.8763E-06  &   4.0730E-07  &   7.6323E-07  &   -4.9469E-04 &   3.6358E-04  &   4.4681E-07  &   3.8120E-07  &    -1.8662E-01    &   1.5722E-02 \\
F390N   &   3.5232E-04  &   6.9202E-04  &   7.0536E-07  &   2.0735E-07  &   -6.5044E-04 &   4.2021E-04  &   5.6012E-07  &   5.7743E-07  &    -1.3178E-01    &   4.7611E-02  \\
F410M   &   6.5712E-04  &   4.2224E-05  &   2.5601E-07  &   7.3600E-07  &   -4.7664E-04 &   3.4715E-04  &   3.9812E-07  &   4.2740E-07  &    -2.1619E-01    &   1.9315E-03 \\
F437N   &   8.1889E-04  &   2.3125E-04  &   5.8299E-08  &   1.1988E-06  &   -2.7334E-04 &   6.4360E-05  &   1.3686E-07  &   1.0929E-08  &    -2.7103E-01    &   1.0515E-01  \\
F439W   &   7.7747E-04  &   5.5030E-04  &   6.9320E-08  &   4.4914E-08  &   -1.9938E-04 &   8.8656E-05  &   7.2636E-08  &   1.4061E-08  &    -2.8276E-01    &   3.2726E-02 \\
F450W   &   1.0615E-03  &   4.5297E-04  &   -2.8400E-07 &   1.7072E-07  &   3.7253E-05  &   1.7591E-04  &   -1.1473E-07 &   2.2917E-07  &    -3.9228E-01    &   2.7343E-02 \\
F467M   &   6.4784E-04  &   8.7865E-05  &   2.2472E-07  &   7.4005E-07  &   -3.9988E-04 &   3.1323E-04  &   4.0137E-07  &   3.2993E-07  &    -2.5186E-01    &   8.9697E-02  \\
F469N   &   7.3305E-04  &   8.2430E-04  &   1.2108E-07  &   5.8161E-07  &   -3.6704E-04 &   7.9614E-05  &   4.1000E-07  &   2.8625E-07  &    -2.7644E-01    &   4.6432E-02 \\
F487N   &   7.4169E-04  &   6.0302E-05  &   2.5542E-08  &   6.8826E-07  &   -4.3557E-04 &   9.3405E-05  &   4.2148E-07  &   1.5703E-07  &    -2.5381E-01    &   4.5522E-02  \\
F502N   &   8.2839E-04  &   3.8789E-04  &   -3.9142E-08 &   3.8289E-07  &   6.6635E-05  &   4.8688E-04  &   -1.3179E-07 &   6.9152E-07  &    -3.6676E-01    &   6.9344E-02  \\
F547M   &   6.2765E-04  &   1.2854E-03  &   1.6569E-07  &   1.2710E-06  &   -4.0714E-04 &   8.5240E-04  &   4.3066E-07  &   1.0717E-06  &    -2.4482E-01    &   5.5599E-02  \\
F555W   &   5.1904E-04  &   6.4443E-04  &   3.1286E-07  &   1.9077E-07  &   -4.7680E-04 &   2.7286E-04  &   5.0747E-07  &   4.4914E-07  &    -2.2501E-01    &   9.2085E-03  \\
F569W   &   4.2361E-04  &   6.9952E-04  &   4.3308E-07  &   9.7134E-08  &   -5.2297E-04 &   1.5238E-04  &   5.9085E-07  &   2.5122E-07  &    -2.0444E-01    &   3.7330E-02 \\
F588N   &   3.4997E-05  &   5.7783E-04  &   9.7718E-07  &   6.8676E-08  &   -9.1938E-04 &   3.0047E-05  &   1.2175E-06  &   1.1055E-07  &    -9.8822E-02    &   1.0007E-02 \\
F606W   &   3.1338E-04  &   5.1021E-04  &   5.5736E-07  &   5.3479E-08  &   -6.2644E-04 &   3.6661E-04  &   7.6919E-07  &   4.4662E-07  &    -1.7611E-01    &   5.5240E-02 \\
F622W   &   3.7037E-04  &   3.5250E-04  &   5.0154E-07  &   1.3738E-07  &   -6.3472E-04 &   2.5299E-05  &   7.9286E-07  &   2.2994E-08  &    -1.7727E-01    &   3.0938E-02  \\
F631N   &   1.2988E-04  &   5.1113E-04  &   8.7971E-07  &   5.2128E-08  &   -7.4011E-04 &   5.8011E-04  &   1.0353E-06  &   6.9121E-07  &    -1.5630E-01    &   8.6977E-02  \\
F656N   &   2.3608E-04  &   7.0643E-04  &   5.9122E-07  &   3.6564E-07  &   -9.1436E-04 &   3.1803E-04  &   1.2730E-06  &   5.6337E-07  &    -9.6546E-02    &   5.0197E-02 \\
F658N   &   3.3598E-04  &   1.2358E-03  &   4.5565E-07  &   1.0107E-06  &   -8.2374E-04 &   3.4227E-04  &   1.1504E-06  &   4.3671E-07  &    -1.2077E-01    &   1.3843E-02 \\
F673N   &   4.5249E-04  &   9.7571E-04  &   3.2815E-07  &   7.0874E-07  &   -7.3098E-04 &   3.8033E-04  &   1.0266E-06  &   5.9859E-07  &    -1.5644E-01    &   9.3436E-02 \\
F675W   &   3.1930E-04  &   1.4704E-03  &   5.2933E-07  &   1.6971E-06  &   -6.6012E-04 &   4.9301E-04  &   8.9662E-07  &   6.8212E-07  &    -1.4834E-01    &   1.5265E-01 \\
F702W   &   4.0199E-04  &   6.9676E-04  &   4.0229E-07  &   4.2661E-07  &   -6.8145E-04 &   3.7542E-04  &   9.1335E-07  &   4.8555E-07  &    -1.4737E-01    &   3.2376E-02  \\
F785LP  &   3.5440E-04  &   1.6001E-04  &   4.2513E-07  &   4.6483E-07  &   -5.1789E-04 &   1.7815E-04  &   6.9719E-07  &   9.3942E-09  &    -8.8608E-02    &   1.7197E-02 \\
F791W   &   4.2940E-04  &   1.9837E-05  &   2.9168E-07  &   6.7819E-07  &   -6.3617E-04 &   1.8034E-04  &   8.4326E-07  &   3.0479E-07  &    -9.7881E-02    &   4.8085E-02  \\
F814W   &   4.9351E-04  &   1.2680E-03  &   2.1637E-07  &   1.2785E-06  &   -5.5825E-04 &   3.4614E-04  &   7.4751E-07  &   6.2545E-07  &    -1.1977E-01    &   3.9957E-02 \\
F850LP  &   4.3779E-04  &   3.2418E-04  &   3.2668E-07  &   3.7159E-07  &   -9.1657E-04 &   6.5723E-04  &   1.1844E-06  &   8.8459E-07  &    -2.7969E-02    &   9.7765E-02  \\
F953N   &   3.5453E-04  &   6.1029E-04  &   3.6232E-07  &   9.9314E-08  &   -8.6414E-04 &   2.4929E-05  &   1.1964E-06  &   4.9795E-08  &    -2.9005E-02    &   1.8024E-02 \\
F1042M  &   3.6873E-04  &   6.9717E-04  &   3.3098E-07  &   3.1197E-07  &   -5.9404E-04 &   2.2777E-04  &   7.8512E-07  &   3.0341E-07  &    -4.8895E-02    &   3.5124E-02 \\
\enddata

\end{deluxetable}

\clearpage

\begin{deluxetable}{cccccccccc}
\tabletypesize{\scriptsize} \rotate \tablecaption{Transformation
Coefficients for differences in Y for CCD 2 of the
WFPC2.\label{tab:table6}}

\tablewidth{0pt}

\tablehead{ \colhead{Filter} & \colhead{AY} & \colhead{$\Delta$
AY} & \colhead{BY} & \colhead{$\Delta$ BY} & \colhead{CY} &
\colhead{$\Delta$ CY} & \colhead{EY} & \colhead{$\Delta$ EY} }

\startdata
F122M   &   -1.2769E-03 &   7.2116E-04  &   1.7019E-06  &   1.4061E-07  &   2.3447E-04  &   3.8577E-05  &   1.3313E-01  &    2.6870E-02  \\
F130LP  &   -5.9736E-04 &   1.4958E-04  &   6.2695E-07  &   5.0544E-07  &   6.2741E-04  &   1.1215E-04  &   -2.5143E-01 &    8.4180E-02  \\
F157W   &   -5.7413E-04 &   1.9765E-04  &   9.2959E-07  &   9.2729E-07  &   1.2856E-04  &   1.7082E-05  &   4.9266E-02  &    6.3530E-02  \\
F160BW  &   -5.9787E-04 &   6.8721E-04  &   9.5391E-07  &   1.5350E-07  &   4.9549E-05  &   5.0500E-05  &   8.0458E-02  &    4.9461E-03  \\
F165LP  &   -5.8566E-04 &   8.2475E-04  &   6.1998E-07  &   5.0210E-07  &   6.4251E-04  &   5.2561E-05  &   -2.5934E-01 &    3.3988E-02  \\
F170W   &   -4.8912E-04 &   1.5294E-04  &   5.8394E-07  &   9.6370E-07  &   4.0667E-04  &   9.4049E-05  &   -7.8870E-02 &    1.9869E-02  \\
F185W   &   -3.2368E-04 &   1.4289E-05  &   2.9727E-07  &   1.0270E-06  &   6.5653E-04  &   1.2303E-05  &   -3.0500E-01 &    5.8139E-02  \\
F218W   &   -5.9047E-04 &   7.5792E-04  &   5.9242E-07  &   3.6696E-07  &   7.5231E-04  &   2.4506E-05  &   -3.3221E-01 &    1.8793E-02  \\
F255W   &   -7.0861E-04 &   5.7714E-05  &   9.2858E-07  &   8.8376E-07  &   6.6426E-04  &   2.4592E-05  &   -2.6962E-01 &    6.9419E-02  \\
F300W   &   -6.9174E-04 &   3.3820E-04  &   5.0603E-07  &   3.0209E-07  &   6.5288E-04  &   1.7783E-04  &   -1.9297E-01 &    9.0924E-02  \\
F336W   &   -9.7424E-04 &   1.6974E-04  &   8.3654E-07  &   5.9423E-07  &   7.0469E-04  &   4.5565E-06  &   -1.5730E-01 &    3.5445E-02  \\
F343N   &   -6.8976E-04 &   2.8786E-04  &   6.0230E-07  &   3.5348E-07  &   6.0548E-04  &   2.4192E-05  &   -1.7345E-01 &    1.7948E-02  \\
F375N   &   -4.6090E-04 &   4.4873E-05  &   2.4345E-07  &   7.2452E-07  &   7.5091E-04  &   1.1604E-04  &   -2.5763E-01 &    1.5506E-03  \\
F380W   &   -7.9694E-04 &   7.3083E-06  &   8.1112E-07  &   7.7081E-07  &   8.1766E-04  &   9.6760E-05  &   -2.8126E-01 &    1.7478E-02  \\
F390N   &   -7.9629E-04 &   6.8864E-04  &   7.5255E-07  &   2.0001E-07  &   7.8171E-04  &   1.8362E-05  &   -2.6068E-01 &    1.0061E-03  \\
F410M   &   -5.3009E-04 &   4.3441E-05  &   4.8386E-07  &   7.3667E-07  &   8.2793E-04  &   4.7060E-05  &   -3.1511E-01 &    3.5278E-02  \\
F437N   &   -3.5790E-04 &   2.3152E-04  &   3.6950E-07  &   1.2025E-06  &   7.7245E-04  &   7.4900E-05  &   -3.4363E-01 &    1.0735E-01  \\
F439W   &   -5.1136E-04 &   5.5162E-04  &   4.7410E-07  &   4.4300E-08  &   8.0554E-04  &   7.8348E-05  &   -3.2021E-01 &    3.1268E-02  \\
F450W   &   -3.0770E-04 &   4.5148E-04  &   2.9770E-07  &   1.7659E-07  &   7.9107E-04  &   1.4936E-05  &   -3.8198E-01 &    7.0102E-03  \\
F467M   &   -3.6886E-04 &   8.8639E-05  &   3.7147E-07  &   7.4238E-07  &   7.6084E-04  &   8.4878E-05  &   -3.5185E-01 &    6.2221E-02  \\
F469N   &   -1.3344E-04 &   8.2657E-04  &   6.0816E-08  &   5.8127E-07  &   7.9325E-04  &   1.1981E-04  &   -3.9619E-01 &    7.0936E-02  \\
F487N   &   -4.2400E-04 &   6.0261E-05  &   4.6738E-07  &   6.9047E-07  &   6.5506E-04  &   1.5637E-05  &   -3.2349E-01 &    3.2425E-02  \\
F502N   &   -5.4847E-04 &   3.9147E-04  &   6.1839E-07  &   3.8016E-07  &   7.5399E-04  &   4.7282E-06  &   -3.3304E-01 &    1.0215E-02  \\
F547M   &   -3.2727E-04 &   1.2929E-03  &   3.2662E-07  &   1.2809E-06  &   7.0405E-04  &   1.0448E-04  &   -3.4899E-01 &    3.6723E-02  \\
F555W   &   -5.2058E-04 &   6.4738E-04  &   5.7756E-07  &   1.9233E-07  &   7.0413E-04  &   4.0157E-05  &   -3.2833E-01 &    2.9269E-02  \\
F569W   &   -4.5306E-04 &   7.0349E-04  &   4.9058E-07  &   1.0103E-07  &   6.9873E-04  &   2.3541E-05  &   -3.3051E-01 &    1.5936E-02  \\
F588N   &   -8.3245E-04 &   5.8144E-04  &   1.0113E-06  &   7.4294E-08  &   6.0344E-04  &   4.6989E-05  &   -2.1832E-01 &    1.9710E-02  \\
F606W   &   -4.9849E-04 &   5.1146E-04  &   5.5429E-07  &   5.3756E-08  &   6.6255E-04  &   5.4921E-05  &   -2.9636E-01 &    1.7063E-02  \\
F622W   &   -6.3191E-04 &   3.5420E-04  &   6.7957E-07  &   1.3465E-07  &   5.9370E-04  &   9.4707E-06  &   -2.2452E-01 &    2.8930E-02  \\
F631N   &   -6.3149E-04 &   5.0943E-04  &   7.5387E-07  &   5.4830E-08  &   6.9639E-04  &   9.8891E-05  &   -2.7664E-01 &    2.8545E-02  \\
F656N   &   -9.1848E-04 &   7.0767E-04  &   1.1068E-06  &   3.6665E-07  &   5.4198E-04  &   7.5165E-05  &   -1.5692E-01 &    2.1607E-03  \\
F658N   &   -7.5737E-04 &   1.2359E-03  &   8.8078E-07  &   1.0102E-06  &   5.3255E-04  &   3.7153E-05  &   -1.7303E-01 &    5.1360E-02  \\
F673N   &   -6.1611E-04 &   9.7439E-04  &   7.1076E-07  &   7.0527E-07  &   5.9861E-04  &   3.6831E-05  &   -2.1955E-01 &    4.2348E-02  \\
F675W   &   -6.4800E-04 &   1.4699E-03  &   7.2206E-07  &   1.6965E-06  &   5.8937E-04  &   1.6853E-05  &   -2.0352E-01 &    9.4255E-02  \\
F702W   &   -6.1372E-04 &   6.9836E-04  &   6.8402E-07  &   4.2820E-07  &   5.8034E-04  &   3.6452E-05  &   -2.0699E-01 &    9.4689E-03  \\
F785LP  &   -6.2499E-04 &   1.2733E-03  &   8.3086E-07  &   4.6363E-07  &   5.3564E-04  &   1.7049E-04  &   -1.7410E-01 &    1.6033E-02  \\
F791W   &   -5.7206E-04 &   3.2595E-04  &   6.8237E-07  &   6.8372E-07  &   5.4464E-04  &   3.2890E-05  &   -1.8338E-01 &    7.4639E-02  \\
F814W   &   -5.5299E-04 &   6.0857E-04  &   6.5498E-07  &   1.2864E-06  &   5.3780E-04  &   9.0472E-05  &   -1.8409E-01 &    9.4117E-02  \\
F850LP  &   -5.1040E-04 &   6.9704E-04  &   6.3386E-07  &   3.7098E-07  &   5.1046E-04  &   4.1144E-05  &   -1.7575E-01 &    2.2404E-02  \\
F953N   &   -3.4970E-04 &   6.9704E-04  &   4.6885E-07  &   9.5477E-08  &   5.0572E-04  &   1.0057E-05  &   -2.0409E-01 &    2.2273E-02  \\
F1042M  &   -7.3933E-04 &   6.9704E-04  &   8.5158E-07  &   3.1111E-07  &   5.7918E-04  &   1.5980E-05  &   -1.5354E-01 &    9.0376E-03  \\
\enddata

\end{deluxetable}

\clearpage

\begin{deluxetable}{cccccccccccc}
\tabletypesize{\tiny} \rotate \tablecaption{Transformation
Coefficients for differences in X for CCDs 1, 3 and 4 of the
WFPC2.\label{tab:table7}}

\tablewidth{0pt}

\tablehead{ \colhead{CCD}  & \colhead{Filter} & \colhead{AX} &
\colhead{$\Delta$ AX} & \colhead{BX} & \colhead{$\Delta$ BX} &
\colhead{CX} & \colhead{$\Delta$ CX} & \colhead{DX}   &
\colhead{$\Delta$ DX}     &\colhead{EX} & \colhead{$\Delta$ EX} }

\startdata
1   &   F336W   &   -1.4712E-03 &   2.0314E-04  &   1.9314E-06  &   3.0081E-07  &   5.4538E-04  &   1.0603E-04  &   -9.5212E-07 &   6.4987E-08  &    -1.9362E-02    &   6.0546E-02  \\
1   &   F410M   &   -1.4432E-03 &   4.6258E-04  &   2.3208E-06  &   6.3272E-07  &   -4.3297E-04 &   4.8043E-04  &   3.5879E-07  &   6.2355E-07  &    3.5333E-02 &   2.2311E-01  \\
1   &   F439W   &   -1.7248E-03 &   5.3508E-05  &   2.4273E-06  &   1.3141E-07  &   -2.1059E-04 &   2.9190E-04  &   -1.5635E-07 &   3.0600E-07  &    1.3528E-01 &   1.1639E-01  \\
1   &   F502N   &   7.8347E-04  &   3.0844E-04  &   -6.5356E-07 &   4.5991E-07  &   1.5067E-05  &   6.2303E-04  &   -3.7495E-07 &   6.5027E-07  &    -3.6371E-01    &   2.4645E-01  \\
1   &   F547M   &   -1.5683E-03 &   8.8274E-04  &   2.5469E-06  &   1.4006E-06  &   2.7682E-04  &   1.5131E-03  &   -6.1970E-07 &   1.7151E-06  &    -6.5746E-02    &   4.7196E-01  \\
1   &   F631N   &   -2.5370E-04 &   5.1158E-04  &   9.0986E-07  &   7.0337E-07  &   4.8444E-04  &   2.3458E-04  &   -8.6157E-07 &   2.8542E-07  &    -2.8728E-01    &   1.0637E-01  \\
1   &   F656N   &   -4.5701E-04 &   5.5195E-04  &   1.4910E-06  &   8.8638E-07  &   -9.8651E-04 &   5.4451E-04  &   7.9465E-07  &   5.9978E-07  &    -2.2444E-02    &   2.2998E-01  \\
1   &   F658N   &   -1.0321E-03 &   3.0401E-04  &   2.0749E-06  &   4.9714E-07  &   -1.1692E-03 &   3.8591E-04  &   1.2424E-06  &   3.8924E-07  &    8.4609E-02 &   1.8804E-01  \\
1   &   F673N   &   9.0750E-05  &   8.6795E-06  &   1.1045E-07  &   6.6407E-08  &   1.6215E-03  &   3.0294E-04  &   -1.6496E-06 &   4.2688E-07  &    -5.6356E-01    &   9.4127E-02  \\
1   &   F791W   &   -4.7231E-04 &   7.0829E-06  &   1.6424E-06  &   1.6725E-07  &   -9.6555E-04 &   1.2801E-06  &   3.6608E-07  &   1.1428E-08  &    6.5062E-03 &   3.5230E-02  \\
1   &   F814W   &   -7.6388E-04 &   1.3690E-04  &   1.6464E-06  &   2.4253E-07  &   -9.5579E-04 &   3.4783E-04  &   8.9782E-07  &   4.3784E-07  &    2.5946E-02 &   5.6355E-03  \\
1   &   F953N   &   5.5270E-04  &   4.7218E-04  &   -8.4714E-07 &   5.0096E-07  &   -1.0638E-03 &   8.0301E-04  &   1.2633E-06  &   9.2087E-07  &    -5.2194E-02    &   1.2576E-01  \\
    &       &       &       &       &       &       &       &       &       &       &       \\
3   &   F336W   &   -9.3903E-04 &   1.7714E-05  &   1.1203E-06  &   6.0871E-08  &   -7.1088E-04 &   1.6578E-05  &   9.2439E-07  &   9.9671E-09  &    3.8285E-01 &   5.1883E-02  \\
3   &   F439W   &   -9.2094E-04 &   2.8460E-04  &   1.0598E-06  &   2.4331E-07  &   -4.8950E-04 &   1.1151E-04  &   7.0139E-07  &   1.1284E-07  &    3.6126E-01 &   7.4641E-02  \\
3   &   F502N   &   -4.6865E-04 &   1.9957E-04  &   3.0728E-07  &   1.7794E-07  &   -5.4480E-04 &   2.2385E-04  &   7.1411E-07  &   2.3242E-07  &    3.3409E-01 &   2.6894E-02  \\
3   &   F547M   &   -5.8309E-04 &   2.3344E-04  &   4.4568E-07  &   2.3728E-07  &   -5.3679E-04 &   1.9270E-04  &   7.2014E-07  &   1.8530E-07  &    3.5438E-01 &   3.3480E-02  \\
3   &   F631N   &   -6.2978E-04 &   8.8066E-05  &   3.5894E-07  &   8.3967E-08  &   -4.6601E-04 &   1.7609E-04  &   5.9419E-07  &   1.8479E-07  &    3.4695E-01 &   5.1377E-03  \\
3   &   F656N   &   -5.8761E-04 &   1.5582E-04  &   2.4465E-07  &   1.6267E-07  &   -4.3263E-04 &   1.4691E-04  &   5.6765E-07  &   1.3980E-07  &    3.4016E-01 &   1.7967E-02  \\
3   &   F658N   &   -5.7825E-04 &   1.5527E-04  &   2.6301E-07  &   1.6838E-07  &   -4.9245E-04 &   1.6775E-04  &   6.4104E-07  &   1.6163E-07  &    3.4166E-01 &   1.2507E-02  \\
3   &   F673N   &   -5.7242E-04 &   1.1241E-04  &   2.9280E-07  &   1.2136E-07  &   -5.2409E-04 &   1.6485E-04  &   6.9031E-07  &   1.7423E-07  &    3.3714E-01 &   7.3933E-03  \\
3   &   F791W   &   -7.8601E-04 &   1.1775E-04  &   4.1279E-07  &   1.0035E-07  &   -6.0751E-04 &   6.6850E-05  &   8.8784E-07  &   3.1242E-08  &    4.1552E-01 &   2.3553E-02  \\
3   &   F814W   &   -7.2630E-04 &   1.3102E-04  &   3.5254E-07  &   1.1169E-07  &   -5.5345E-04 &   6.5002E-05  &   8.1440E-07  &   3.7073E-08  &    3.9633E-01 &   2.7509E-02  \\
3   &   F953N   &   -9.2573E-04 &   5.3979E-05  &   5.8281E-07  &   7.2124E-08  &   -3.0597E-04 &   3.8342E-05  &   3.2773E-07  &   2.5017E-08  &    3.8498E-01 &   9.2880E-03  \\
    &       &       &       &       &       &       &       &       &       &       & \\
4   &   F336W   &   -7.0842E-05 &   6.0241E-05  &   1.0991E-07  &   1.2679E-07  &   5.4728E-05  &   7.6084E-05  &   5.7422E-08  &   4.2918E-08  &    8.0641E-02 &   7.4788E-03  \\
4   &   F410M   &   1.6924E-04  &   4.1626E-05  &   -1.5847E-07 &   1.3140E-07  &   2.7139E-04  &   1.3844E-04  &   -1.9485E-07 &   1.2966E-07  &    1.8544E-03 &   7.6400E-03  \\
4   &   F439W   &   -1.0620E-03 &   2.8368E-04  &   6.2002E-07  &   3.4300E-07  &   4.1942E-05  &   1.5227E-04  &   6.7529E-08  &   1.9327E-07  &    -2.0855E-01    &   5.4459E-03  \\
4   &   F502N   &   -5.1309E-05 &   1.7614E-04  &   3.1372E-08  &   1.4909E-07  &   1.0392E-04  &   8.7079E-05  &   -2.3511E-08 &   6.8692E-08  &    6.4091E-02 &   4.7144E-02  \\
4   &   F547M   &   -4.4688E-05 &   6.9297E-05  &   -7.7930E-08 &   3.2813E-08  &   2.1815E-04  &   7.5754E-05  &   -6.8311E-08 &   5.6401E-08  &    6.8528E-02 &   2.6194E-02  \\
4   &   F631N   &   2.6080E-04  &   9.4397E-05  &   -4.7310E-07 &   1.1434E-07  &   5.9569E-04  &   5.3373E-06  &   -5.1738E-07 &   1.5702E-08  &    -7.9896E-02    &   3.1789E-02  \\
4   &   F656N   &   5.5046E-05  &   7.4871E-05  &   -2.2209E-07 &   7.3680E-08  &   4.9425E-04  &   3.8794E-05  &   -3.3328E-07 &   4.1472E-08  &    -5.6229E-02    &   2.3318E-02  \\
4   &   F658N   &   8.9618E-06  &   6.4151E-05  &   -1.8333E-07 &   5.8458E-08  &   3.2882E-04  &   2.3197E-05  &   -1.5867E-07 &   2.1859E-08  &    -1.2142E-02    &   2.4446E-02  \\
4   &   F673N   &   -7.3280E-06 &   8.4427E-05  &   -1.5732E-07 &   8.0974E-08  &   3.9325E-04  &   3.6719E-05  &   -2.1421E-07 &   3.8549E-08  &    -1.9680E-02    &   2.4815E-02  \\
4   &   F791W   &   -4.7632E-05 &   1.0345E-05  &   -2.7834E-07 &   4.7405E-08  &   1.7045E-04  &   5.4677E-05  &   1.0261E-07  &   1.1569E-07  &    7.6594E-02 &   1.8588E-02  \\
4   &   F814W   &   -1.0220E-04 &   1.3400E-05  &   -2.1858E-07 &   1.3492E-08  &   2.8019E-04  &   1.4567E-06  &   -7.9126E-08 &   3.3726E-08  &    7.6468E-02 &   1.4964E-02  \\
4   &   F953N   &   -1.9506E-04 &   3.2979E-05  &   -4.8535E-08 &   4.3529E-08  &   7.6459E-04  &   5.8537E-06  &   -6.2222E-07 &   5.5928E-09  &    -2.5899E-02    &   2.0817E-02  \\

\enddata

\end{deluxetable}
\clearpage

\begin{deluxetable}{cccccccccc}
\tabletypesize{\scriptsize} \rotate \tablecaption{Transformation
Coefficients for differences in Y for CCDs 1, 3 and 4 of the
WFPC2.\label{tab:table8}}

\tablewidth{0pt}

\tablehead{ \colhead{CCD} & \colhead{Filter} & \colhead{AY} &
\colhead{$\Delta$ AY} & \colhead{BY} & \colhead{$\Delta$ BY} &
\colhead{CY} & \colhead{$\Delta$ CY} & \colhead{EY}   &
\colhead{$\Delta$ EY} }

\startdata
1   &   F336W   &   -4.5267E-04 &   1.7088E-04  &   5.1804E-07  &   1.1211E-07  &   2.4592E-04  &   5.3033E-05  &   -2.2030E-01 &   1.4162E-02 \\
1   &   F410M   &   -1.9323E-03 &   9.2308E-05  &   2.7114E-06  &   1.2272E-07  &   2.5148E-05  &   7.3059E-05  &   -3.5450E-03 &   7.0452E-02 \\
1   &   F439W   &   -1.6081E-03 &   5.3016E-05  &   2.3206E-06  &   8.6138E-08  &   -1.5549E-04 &   1.1260E-05  &   2.0490E-02  &   7.6175E-02 \\
1   &   F502N   &   -1.4251E-04 &   9.2454E-05  &   -2.6831E-08 &   2.2215E-07  &   -4.2667E-05 &   8.7347E-05  &   -2.5410E-01 &   1.2533E-01 \\
1   &   F547M   &   -1.9137E-03 &   6.0862E-04  &   2.6601E-06  &   7.9182E-07  &   3.5608E-05  &   4.6194E-05  &   -5.8677E-02 &   1.6615E-01 \\
1   &   F631N   &   1.0882E-04  &   1.7776E-04  &   -1.9568E-07 &   1.9795E-07  &   1.1280E-04  &   2.0756E-05  &   -4.0566E-01 &   4.1088E-02 \\
1   &   F656N   &   -8.9587E-04 &   4.7854E-04  &   1.6163E-06  &   7.5881E-07  &   -9.3119E-05 &   1.1899E-06  &   -2.5839E-01 &   1.2473E-01 \\
1   &   F658N   &   -8.8573E-04 &   5.0645E-04  &   1.3768E-06  &   7.1060E-07  &   -1.6862E-04 &   3.4322E-05  &   -1.5266E-01 &   1.6439E-01 \\
1   &   F673N   &   -8.7077E-05 &   1.0665E-03  &   2.7405E-07  &   1.2817E-06  &   9.6639E-05  &   7.6852E-05  &   -3.5289E-01 &   2.4351E-01 \\
1   &   F791W   &   -1.4453E-03 &   2.6025E-05  &   1.8092E-06  &   1.9504E-07  &   -1.7679E-04 &   4.0822E-05  &   -1.8141E-02 &   1.0879E-02 \\
1   &   F814W   &   -2.0705E-03 &   8.8579E-05  &   2.1976E-06  &   3.4328E-08  &   2.0374E-04  &   8.3431E-05  &   -7.3640E-04 &   2.7700E-03 \\
1   &   F953N   &   -8.2649E-04 &   7.9845E-05  &   5.2590E-07  &   1.0979E-08  &   2.6318E-04  &   6.4035E-05  &   -2.4755E-01 &   6.2983E-02 \\
    &           &               &               &               &               &               &               &               &              \\
3   &   F336W   &   -2.8360E-04 &   5.1938E-05  &   1.9602E-07  &   7.0531E-08  &   3.4580E-05  &   4.5088E-05  &   1.8913E-01  &   3.3603E-02 \\
3   &   F439W   &   -2.6797E-04 &   8.5947E-05  &   2.4919E-07  &   1.0787E-07  &   -8.1453E-06 &   7.7726E-05  &   2.3756E-01  &   4.9785E-02 \\
3   &   F502N   &   9.5659E-05  &   6.4347E-05  &   -3.0898E-07 &   4.4655E-08  &   -5.7903E-05 &   7.4473E-05  &   2.0546E-01  &   4.3057E-02 \\
3   &   F547M   &   -2.1216E-04 &   1.1559E-04  &   1.4678E-07  &   9.4336E-08  &   -1.6037E-04 &   6.0825E-05  &   2.9655E-01  &   2.5385E-02 \\
3   &   F631N   &   7.2237E-05  &   1.8187E-04  &   -2.6633E-07 &   1.9962E-07  &   -2.1996E-04 &   1.9814E-05  &   2.4502E-01  &   7.3686E-03 \\
3   &   F656N   &   -1.7747E-05 &   1.1005E-04  &   -1.5898E-07 &   1.2703E-07  &   -2.2536E-04 &   3.2498E-05  &   2.5347E-01  &   9.8435E-03 \\
3   &   F658N   &   -1.7834E-05 &   1.5657E-04  &   -1.1650E-07 &   1.7216E-07  &   -2.4005E-04 &   3.7801E-05  &   2.5356E-01  &   4.0406E-03 \\
3   &   F673N   &   1.4971E-04  &   1.6268E-04  &   -2.9553E-07 &   1.9097E-07  &   -2.7553E-04 &   2.3091E-05  &   2.3886E-01  &   2.9042E-03 \\
3   &   F791W   &   -1.2878E-04 &   1.1924E-04  &   1.3474E-07  &   1.0093E-07  &   -3.9627E-04 &   3.9766E-05  &   3.3716E-01  &   7.4384E-03 \\
3   &   F814W   &   -2.9778E-05 &   8.2745E-05  &   1.4673E-08  &   5.4746E-08  &   -3.8251E-04 &   4.7535E-05  &   3.2137E-01  &   1.6489E-02 \\
3   &   F953N   &   1.3593E-04  &   6.5447E-05  &   -3.4619E-07 &   7.2936E-08  &   -3.9784E-04 &   5.8495E-06  &   2.8949E-01  &   1.0085E-03 \\
    &           &               &               &               &               &               &               &               &              \\
4   &   F336W   &   1.3596E-04  &   4.7610E-05  &   -7.1020E-08 &   8.6652E-08  &   1.0208E-05  &   1.8239E-05  &   6.4880E-02  &   2.7470E-02 \\
4   &   F410M   &   1.0364E-04  &   7.2133E-05  &   1.0857E-07  &   5.3161E-08  &   5.8320E-05  &   2.1331E-05  &   3.5865E-02  &   1.1514E-02 \\
4   &   F439W   &   3.1437E-04  &   2.1234E-04  &   3.0137E-07  &   2.4324E-07  &   -5.4500E-06 &   2.6052E-05  &   -8.8196E-01 &   1.0723E-03 \\
4   &   F502N   &   -1.2490E-04 &   5.2234E-05  &   3.4937E-07  &   4.0502E-08  &   -6.6212E-05 &   2.1759E-05  &   1.3330E-01  &   1.5952E-02 \\
4   &   F547M   &   -7.5882E-05 &   1.5468E-04  &   4.1111E-07  &   1.4637E-07  &   -1.2338E-04 &   2.5181E-05  &   1.1845E-01  &   2.3088E-03 \\
4   &   F631N   &   -4.5352E-05 &   1.5893E-04  &   3.4927E-07  &   1.8279E-07  &   -1.9190E-04 &   2.1165E-05  &   1.2784E-01  &   3.6521E-03 \\
4   &   F656N   &   7.7703E-06  &   5.6011E-05  &   2.6099E-07  &   7.0138E-08  &   -1.8331E-04 &   1.2293E-05  &   1.1700E-01  &   9.1849E-03 \\
4   &   F658N   &   1.1524E-04  &   7.0109E-05  &   1.4934E-07  &   8.8199E-08  &   -2.0213E-04 &   1.3322E-05  &   1.0481E-01  &   8.2847E-03 \\
4   &   F673N   &   8.2954E-05  &   9.4264E-05  &   1.7366E-07  &   1.0789E-07  &   -2.2335E-04 &   2.6298E-05  &   1.1249E-01  &   9.7699E-03 \\
4   &   F791W   &   2.1294E-04  &   6.0643E-05  &   2.5708E-08  &   5.5320E-08  &   -2.9623E-04 &   3.0729E-05  &   1.2394E-01  &   1.2264E-02 \\
4   &   F814W   &   1.0914E-04  &   3.2539E-05  &   1.3706E-07  &   3.1141E-08  &   -2.9638E-04 &   2.9801E-05  &   1.4806E-01  &   1.8647E-02 \\
4   &   F953N   &   1.6018E-04  &   6.6380E-06  &   7.2974E-08  &   1.3107E-08  &   -2.6701E-04 &   4.2020E-06  &   9.3355E-02  &   9.5677E-03 \\

\enddata

\end{deluxetable}

\begin{table}[!h]
\caption{Accuracy of the Measuring Algorithm for the CCDs of the WFPC2.} \label{tab:ACCURACY}\centering

\begin{tabular}{|c|c|c|}

\hline
     CCD            &         Accuracy(X) (pixels)     &     Accuracy (Y) (pixels)      \\\hline
                    &                                  &                                \\
     1              &         $\pm 0.09$               &      $\pm 0.13$                \\
     2              &         $\pm 0.03$               &      $\pm 0.02$                \\
     3              &         $\pm 0.06$               &      $\pm 0.05$                \\
     4              &         $\pm 0.03$               &      $\pm 0.02$                \\\hline

\end{tabular}
\end{table}

\end{document}